# Approximate spin and pseudospin solutions of the Dirac equation for the inversely quadratic Yukawa potential and tensor interaction

M. Hamzavi[1*], S. M. Ikhdair[2**], B. I. Ita[3],

[1]*Department of Basic Sciences, Shahrood Branch, Islamic Azad University, Shahrood, Iran*
[2]*Physics Department, Near East University, Nicosia, North Cyprus, Mersin 10, Turkey*
[3]*Theoretical Chemistry Group, Department of Pure & Applied Chemistry, University of Calabar, Nigeria*

[*]*Corresponding author: Tel.:+98 273 3395270, fax: +98 273 3395270*
Email: majid.hamzavi@gmail.com
[**]Email: sikhdair@neu.edu.tr

**Abstract**

We approximately solve the Dirac equation for the inversely quadratic Yukawa (IQY) potential including a Coulomb-like tensor potential with arbitrary spin-orbit coupling quantum number $\kappa$. In the framework of the spin and pseudospin (pspin) symmetry, we obtain the energy eigenvalue equation and the corresponding eigenfunctions in closed form by using the Nikiforov-Uvarov method. The numerical results show that the Coulomb-like tensor interaction removes degeneracies between spin and pspin state doublets.

## 1. Introduction

Relativistic symmetries of the Dirac Hamiltonian had been discovered many years ago. However, these symmetries have been recently recognized empirically in nuclear and hadronic spectroscopies [1]. Within the framework of Dirac equation, the pspin symmetry used to feature deformed nuclei and the superdeformation to establish an effective shell-model [2-4]. The spin symmetry is relevant to mesons [5] and occurs when the difference of the scalar $S(r)$ and vector $V(r)$ potentials are constant, i.e., $\Delta(r) = C_s$ and the pspin symmetry occurs when the sum of the scalar and vector potentials are constant, i.e., $\Sigma(r) = C_{ps}$ [6-7]. The pspin symmetry refers to a quasi-degeneracy of single nucleon doublets with non-relativistic quantum number $(n, l, j = l + 1/2)$ and $(n - 1, l + 2, j = l + 3/2)$, where $n$, $l$ and $j$ denote the single nucleon radial, orbital and total angular quantum numbers, respectively [8-9]. Further, the total angular momentum is $j = \tilde{l} + \tilde{s}$, where $\tilde{l} = l + 1$ is pseudo-angular momentum and $\tilde{s}$ is pspin angular momentum [10]. Recently, the tensor potentials were introduced into the Dirac equation with the substitution $\vec{p} \to \vec{p} - im\omega\beta.\hat{r}U(r)$ and a spin-orbit coupling is added to the Dirac Hamiltonian [11-12]. Lisboa et al. [13] have studied a generalized relativistic harmonic oscillator for spin-1/2 particles by considering a Dirac Hamiltonian that contains quadratic vector and scalar potentials together with a linear tensor potential under the conditions of pspin and spin symmetry. Alberto et al. [14] studied the contribution of the isoscalar tensor coupling to the realization of pspin symmetry in nuclei. Akçay [15] solved exactly the Dirac equation with scalar and vector quadratic potentials including a Coulomb-like tensor potential. He also solved exactly the Dirac equation for a linear and Coulomb-like term containing the tensor potential too [16]. Overmore, Aydoğdu and Sever [17] obtained exact solution of the Dirac equation for the pseudoharmonic potential in the presence of linear tensor potential under the pspin symmetry and showed that tensor interactions remove all degeneracies between members of pspin doublets. Ikhdair and Sever [10] solved Dirac equation approximately for Hulthén potential including Coulomb-like tensor potential with arbitrary spin-orbit coupling number $\kappa$ under spin and pspin symmetry limit. Very recently, Hamzavi et al. [18,19] presented exact solutions of the Dirac equation for Mie-type potential and approximate solutions of the Dirac equation for Morse potential with a Coulomb-like tensor potential. There are many works on the solutions of the Schrödinger, Klein-Gordon (KG) and the Dirac equations for various types of potentials by different authors [20-38].

In the present work, our aim is to solve the Dirac equation for the inversely quadratic Yukawa (IQY) potential in the presence of spin and pspin symmetries and by including a coulomb-like tensor potential. The IQY potential takes the following form

$$V(r) = -\frac{V_0}{r^2} e^{-2\alpha r}, \tag{1}$$

where $\alpha$ is the screening parameter and $V_0$ is the depth of the potential. A form of Yukawa potential has been earlier used by Taseli [39] in obtaining modified Laguerre basis for hydrogen-like systems. Also, Kermode *et al.* [40] have used different forms of the Yukawa potential to obtain the effective range functions. But, not much has been done in solving the IQY potential.

This paper is organized as follows. in Section 2, we briefly introduce the Dirac equation with scalar and vector potential with arbitrary spin-orbit coupling quantum number $\kappa$ including tensor interaction under spin and pspin symmetry limits. The Nikiforov-Uvarov (NU) method is presented in Section 3. The energy eigenvalue equations and corresponding eigenfunctions are obtained in Section 4. In this section, some remarks and numerical results are also presented. Finally, our conclusion is given in Section 5.

## 2. The Dirac equation with tensor coupling potential

The Dirac equation for fermionic massive spin-$1/2$ particles moving in the field of an attractive scalar potential $S(r)$, a repulsive vector potential $V(r)$ and a tensor potential $U(r)$ (in units $\hbar = c = 1$) is

$$[\vec{\alpha}.\vec{p} + \beta(M + S(r)) - i\beta\vec{\alpha}.\hat{r}U(r)]\psi(\vec{r}) = [E - V(r)]\psi(\vec{r}), \qquad (2)$$

where $E$ is the relativistic binding energy of the system, $\vec{p} = -i\vec{\nabla}$ is the three-dimensional momentum operator and $M$ is the mass of the fermionic particle. $\vec{\alpha}$ and $\beta$ are the $4\times 4$ usual Dirac matrices given by

$$\vec{\alpha} = \begin{pmatrix} 0 & \vec{\sigma} \\ \vec{\sigma} & 0 \end{pmatrix}, \quad \beta = \begin{pmatrix} I & 0 \\ 0 & -I \end{pmatrix}, \qquad (3)$$

where $I$ is $2\times 2$ unitary matrix and $\vec{\sigma}$ are three-vector spin matrices

$$\sigma_1 = \begin{pmatrix} 0 & 1 \\ 1 & 0 \end{pmatrix}, \quad \sigma_2 = \begin{pmatrix} 0 & -i \\ i & 0 \end{pmatrix}, \quad \sigma_3 = \begin{pmatrix} 1 & 0 \\ 0 & -1 \end{pmatrix}. \qquad (4)$$

The eigenvalues of spin-orbit coupling operator are $\kappa = \left(j + \frac{1}{2}\right) > 0$ and $\kappa = -\left(j + \frac{1}{2}\right) < 0$ for unaligned spin $j = l - \frac{1}{2}$ and the aligned spin $j = l + \frac{1}{2}$, respectively. The set $(H^2, K, J^2, J_z)$ can be taken as the complete set of the conservative quantities with $\vec{J}$ is the total angular momentum operator and $K = (\vec{\sigma}.\vec{L}+1)$ is the spin-orbit where $\vec{L}$ is orbital angular momentum of the spherical nucleons that commutes with the Dirac Hamiltonian. Thus, the spinor wave functions can be classified according to their angular momentum $j$, the spin-orbit quantum number $\kappa$, and the radial quantum number $n$. Hence, they can be written as follows

$$\psi_{n\kappa}(\vec{r}) = \begin{pmatrix} f_{n\kappa}(\vec{r}) \\ g_{n\kappa}(\vec{r}) \end{pmatrix} = \frac{1}{r}\begin{pmatrix} F_{n\kappa}(r)Y_{jm}^{l}(\theta,\varphi) \\ iG_{n\kappa}(r)Y_{jm}^{\tilde{l}}(\theta,\varphi) \end{pmatrix}, \tag{5}$$

where $f_{n\kappa}(\vec{r})$ is the upper (large) component and $g_{n\kappa}(\vec{r})$ is the lower (small) component of the Dirac spinors. $Y_{jm}^{l}(\theta,\varphi)$ and $Y_{jm}^{\tilde{l}}(\theta,\varphi)$ are spin and pspin spherical harmonics, respectively, and $m$ is the projection of the angular momentum on the $z$-axis. Substituting Eq. (5) into Eq. (2) and making use of the following relations

$$(\vec{\sigma}.\vec{A})(\vec{\sigma}.\vec{B}) = \vec{A}.\vec{B} + i\vec{\sigma}.(\vec{A}\times\vec{B}), \tag{6a}$$

$$(\vec{\sigma}.\vec{P}) = \vec{\sigma}.\hat{r}\left(\hat{r}.\vec{P} + i\frac{\vec{\sigma}.\vec{L}}{r}\right), \tag{6b}$$

together wth the properties

$$\begin{aligned}
(\vec{\sigma}.\vec{L})Y_{jm}^{\tilde{l}}(\theta,\phi) &= (\kappa-1)Y_{jm}^{\tilde{l}}(\theta,\phi), \\
(\vec{\sigma}.\vec{L})Y_{jm}^{l}(\theta,\phi) &= -(\kappa-1)Y_{jm}^{l}(\theta,\phi), \\
(\vec{\sigma}.\hat{r})Y_{jm}^{\tilde{l}}(\theta,\phi) &= -Y_{jm}^{l}(\theta,\phi), \\
(\vec{\sigma}.\hat{r})Y_{jm}^{l}(\theta,\phi) &= -Y_{jm}^{\tilde{l}}(\theta,\phi),
\end{aligned} \tag{7}$$

one obtains two coupled differential equations whose solutions are the upper and lower radial wave functions $F_{n\kappa}(r)$ and $G_{n\kappa}(r)$ as

$$\left(\frac{d}{dr} + \frac{\kappa}{r} - U(r)\right)F_{n\kappa}(r) = (M + E_{n\kappa} - \Delta(r))G_{n\kappa}(r), \tag{8a}$$

$$\left(\frac{d}{dr} - \frac{\kappa}{r} + U(r)\right)G_{n\kappa}(r) = (M - E_{n\kappa} + \Sigma(r))F_{n\kappa}(r), \tag{8b}$$

where

$$\Delta(r) = V(r) - S(r), \tag{9a}$$

$$\Sigma(r) = V(r) + S(r). \tag{9b}$$

After eliminating $F_{n\kappa}(r)$ and $G_{n\kappa}(r)$ from Eqs. (8), we obtain the following two Schrödinger-like differential equations for the upper and lower radial spinor components

$$\left[\frac{d^2}{dr^2}-\frac{\kappa(\kappa+1)}{r^2}+\frac{2\kappa}{r}U(r)-\frac{dU(r)}{dr}-U^2(r)\right]F_{n\kappa}(r)$$
$$+\frac{\frac{d\Delta(r)}{dr}}{M+E_{n\kappa}-\Delta(r)}\left(\frac{d}{dr}+\frac{\kappa}{r}-U(r)\right)F_{n\kappa}(r) \qquad (10)$$
$$=\left[(M+E_{n\kappa}-\Delta(r))(M-E_{n\kappa}+\Sigma(r))\right]F_{n\kappa}(r),$$

$$\left[\frac{d^2}{dr^2}-\frac{\kappa(\kappa-1)}{r^2}+\frac{2\kappa}{r}U(r)+\frac{dU(r)}{dr}-U^2(r)\right]G_{n\kappa}(r)$$
$$+\frac{\frac{d\Sigma(r)}{dr}}{M-E_{n\kappa}+\Sigma(r)}\left(\frac{d}{dr}-\frac{\kappa}{r}+U(r)\right)G_{n\kappa}(r) \qquad (11)$$
$$=\left[(M+E_{n\kappa}-\Delta(r))(M-E_{n\kappa}+\Sigma(r))\right]G_{n\kappa}(r)$$

respectively, where $\kappa(\kappa-1)=\tilde{l}(\tilde{l}+1)$ and $\kappa(\kappa+1)=l(l+1)$. The quantum number $\kappa$ is related to the quantum numbers for spin symmetry $l$ and pspin symmetry $\tilde{l}$ as

$$\kappa=\begin{cases}-(l+1)=-(j+1/2) & (s_{1/2},p_{3/2},etc.)\quad j=l+\frac{1}{2},\ \text{alined spin}\ (\kappa<0)\\ +l=+(j+1/2) & (p_{1/2},d_{3/2},etc.)\quad j=l-\frac{1}{2},\ \text{unalined spin}\ (\kappa>0),\end{cases} \qquad (12)$$

and the quasidegenerate doublet structure can be expressed in terms of a pspin angular momentum $\tilde{s}=1/2$ and pseudo-orbital angular momentum $\tilde{l}$, which is defined as

$$\kappa=\begin{cases}-\tilde{l}=-(j+1/2) & (s_{1/2},p_{3/2},etc.)\quad j=\tilde{l}-\frac{1}{2},\ \text{alined pspin}\ (\kappa<0)\\ +(\tilde{l}+1)=+(j+1/2) & (d_{3/2},f_{5/2},etc.)\quad j=\tilde{l}+\frac{1}{2},\ \text{unalined spin}\ (\kappa>0),\end{cases} \qquad (13)$$

where $\kappa=\pm 1,\pm 2,...$. For example, $(1s_{1/2},0d_{3/2})$ and $(1p_{3/2},0f_{5/2})$ can be considered as pspin doublets.

## 2.1. Pspin symmetry limit

Ginocchio showed that there is a connection between pspin symmetry and near equality of the time component of a vector potential and the scalar potential, $V(r) \approx -S(r)$ [7]. After that, Meng *et al.* derived that if $\frac{d[V(r)+S(r)]}{dr} = \frac{d\Sigma(r)}{dr} = 0$ or $\Sigma(r) = C_{ps} = $ constant, then pspin symmetry is exact in the Dirac equation [41-42]. In this part, we are taking $\Delta(r)$ as IQY potential (1) and tensor potential as Coulomb-like potential, that is,

$$\Delta(r) = -\frac{V_0}{r^2} e^{-2\alpha r}, \tag{14}$$

$$U(r) = -\frac{H}{r}, \quad H = \frac{Z_a Z_b e^2}{4\pi\varepsilon_0}, \quad r \geq R_c, \tag{15}$$

where $R_c = 7.78 fm$ is the Coulomb radius, $Z_a$ and $Z_b$ denote the charges of the projectile $a$ and the target nuclei $b$, respectively [10]. Under this symmetry, Eq. (11), recasts in the simple form:

$$\left[\frac{d^2}{dr^2} - \frac{\kappa(\kappa-1)}{r^2} - \frac{2\kappa H}{r^2} + \frac{H}{r^2} - \frac{H^2}{r^2}\right] G_{n\kappa}(r) = \left[\tilde{\gamma}\left(-\frac{V_0}{r^2}e^{-2\alpha r}\right) + \tilde{\beta}^2\right] G_{n\kappa}(r), \tag{16}$$

where $\kappa = -\tilde{l}$ and $\kappa = \tilde{l}+1$ for $\kappa < 0$ and $\kappa > 0$, respectively. Also, we identified $\tilde{\gamma} = E_{n\kappa} - M - C_{ps}$ and $\tilde{\beta}^2 = (M + E_{n\kappa})(M - E_{n\kappa} + C_{ps})$.

**2.2. Spin symmetry limit**

In the spin symmetry limit $\frac{d\Delta(r)}{dr} = 0$ or $\Delta(r) = C_s = $ constant [41-42], with $\Sigma(r)$ is taken as IQY potential (1) and Coulomb-like potential tensor potential. Thus, Eq. (10) recasts in the form:

$$\left[\frac{d^2}{dr^2} - \frac{\kappa(\kappa+1)}{r^2} - \frac{2\kappa H}{r^2} - \frac{H}{r^2} - \frac{H^2}{r^2}\right] F_{n\kappa}(r) = \left[\gamma\left(-\frac{V_0}{r^2}e^{-2\alpha r}\right) + \beta^2\right] F_{n\kappa}(r), \tag{17}$$

where $\kappa = l$ and $\kappa = -l-1$ for $\kappa < 0$ and $\kappa > 0$, respectively. Also, $\gamma = M + E_{n\kappa} - C_s$ and $\beta^2 = (M - E_{n\kappa})(M + E_{n\kappa} - C_s)$.

Since the Dirac equation with the IQY potential has no exact solution, we use an approximation for the centrifugal term as [43-46]

$$\frac{1}{r^2} \approx 4a^2 \frac{e^{-2ar}}{(1-e^{-2ar})^2} \tag{18}$$

Finally, for the solutions of Eq. (16) and Eq. (17) with the above approximation, we will employ the Nikiforov-Uvarov method which is briefly introduced in the following section.

## 3. The NU method

This method can be used to solve the second order differential equations with an appropriate coordinate transformation $s = s(r)$ [47]

$$\psi_n''(s) + \frac{\tilde{\tau}(s)}{\sigma(s)}\psi_n'(s) + \frac{\tilde{\sigma}(s)}{\sigma^2(s)}\psi_n(s) = 0, \tag{19}$$

where $\sigma(s)$ and $\tilde{\sigma}(s)$ are polynomials, at most of second degree, and $\tilde{\tau}(s)$ is a first-degree polynomial. A solution of Eq. (19) is found by a separation of variables, using the transformation $\psi_n(s) = \phi(s)y_n(s)$. It reduces (19) into an equation of the hypergeometric type

$$\sigma(s)y_n''(s) + \tau(s)y_n'(s) + \lambda y_n(s) = 0. \tag{20}$$

$y_n(s)$ is the hypergeometric-type function whose polynomial solutions are given by Rodrigues relation:

$$y_n(s) = \frac{B_n}{\rho(s)}\frac{d^n}{ds^n}[\sigma^n(s)\rho(s)], \tag{21}$$

where $B_n$ is the normalization constant and the weight function $\rho(s)$ must satisfy the condition [47]

$$\frac{d}{ds}w(s) = \frac{\tau(s)}{\sigma(s)}w(s), \qquad w(s) = \sigma(s)\rho(s). \tag{22}$$

and $\phi(s)$ is defined from its logarithmic derivative relation

$$\frac{\phi'(s)}{\phi(s)} = \frac{\pi(s)}{\sigma(s)}. \tag{23}$$

The function $\pi(s)$ and the parameter $\lambda$, required for this method, are defined as follows

$$\pi(s) = \frac{\sigma' - \tilde{\tau}}{2} \pm \sqrt{\left(\frac{\sigma' - \tilde{\tau}}{2}\right)^2 - \tilde{\sigma} + k\sigma}, \tag{24a}$$

$$\lambda = k + \pi'(s). \tag{24b}$$

In order to find the value of $k$, the expression under the square root must be a square of a polynomial. Thus, a new eigenvalue equation is

$$\lambda = \lambda_n = -n\tau' - \frac{n(n-1)}{2}\sigma'',\qquad(25)$$

where

$$\tau(s) = \tilde{\tau}(s) + 2\pi(s),\qquad(26)$$

and its derivative must be negative [47]. In this regard, one can derive the parametric generalization version of the NU method [48,49] outlined in some details in Appendix A.

## 4. Solutions of the Dirac equation

We are now going to solve the Dirac equation with IQY potential with tensor potential by using the Nikiforov-Uvarov method.

### 4.1. Pspin symmetric case

To obtain solution of Eq. (16), by using transformation $s = e^{-2\alpha r}$, we rewrite it as follows

$$\left[\frac{d^2}{ds^2} + \frac{1-s}{s(1-s)}\frac{d}{ds} + \frac{1}{s^2(1-s)^2}\left(-\Lambda_\kappa(\Lambda_\kappa - 1)s + \tilde{\mathcal{W}}_0 s^2 - \frac{\tilde{\beta}}{4\alpha^2}(1-s)^2\right)\right]G_{n\kappa} = 0,\qquad(27)$$

where $\Lambda_\kappa = \kappa + H$. Comparing Eq. (27) with (A1), we get

$$\begin{aligned}
&\alpha_1 = 1, &&\xi_1 = \frac{\tilde{\beta}}{4\alpha^2} - \tilde{\mathcal{W}}_0 \\
&\alpha_2 = 1, &&\xi_2 = -\Lambda_\kappa(\Lambda_\kappa - 1) + \frac{2\tilde{\beta}}{4\alpha^2} \\
&\alpha_3 = 1, &&\xi_3 = \frac{\tilde{\beta}}{4\alpha^2}
\end{aligned}\qquad(28)$$

and, form (A6)-(A12), we further obtain

$$\begin{aligned}
&\alpha_4 = 0, &&\alpha_5 = -\frac{1}{2} \\
&\alpha_6 = \frac{1}{4} + \frac{\tilde{\beta}}{4\alpha^2} - \tilde{\gamma}V_0, &&\alpha_7 = \Lambda_\kappa(\Lambda_\kappa - 1) - \frac{2\tilde{\beta}}{4\alpha^2} \\
&\alpha_8 = \frac{\tilde{\beta}}{4\alpha^2}, &&\alpha_9 = (\Lambda_\kappa - \frac{1}{2})^2 - \tilde{\gamma}V_0,
\end{aligned}\qquad(29)$$

In addition, the energy eigenvalue equation can be obtained by using the relation (A17) as follows

$$\left(n+\frac{1}{2}+\sqrt{(\Lambda_\kappa-\frac{1}{2})^2-\tilde{\gamma}V_0}+\sqrt{\frac{\tilde{\beta}^2}{4\alpha^2}}\right)^2=\frac{\tilde{\beta}^2}{4\alpha^2}-\tilde{\gamma}V_0. \tag{30}$$

By substituting the explicit forms of $\tilde{\gamma}$ and $\tilde{\beta}^2$ after Eq. (16) into Eq. (30), one can readily obtain closed form for the energy formula. In the limiting case when the screening parameter $\alpha \to 0$ (low screening regime), the potential approximates as $V_{IQY}(r)=-V_0\lim_{\alpha\to 0}\frac{e^{-2\alpha r}}{r^2}\simeq\frac{A}{r^2}-\frac{B}{r}+C$, where the potential parameters are: defined as $A=-V_0$, $B=-2\alpha V_0$, $C=-2\alpha^2 V_0$. This potential is well known as Mie-type potential [18, 27]. The energy eigenvalue equation for this potential has recently been found in Ref. [27] as

$$\sqrt{(E_{n\kappa}-M-C_{ps})C+(M+E_{n\kappa})(M-E_{n\kappa}+C_{ps})}$$

$$=\frac{(E_{n\kappa}-M-C_{ps})B}{1+2n+2\sqrt{\left(\kappa-\frac{1}{2}\right)^2+(E_{n\kappa}-M-C_{ps})A}}. \tag{31}$$

The special case when $A=C=0$ and $C_{ps}=0$ yields the energy formula for the Coulomb-like potential, i.e. [27,50]

$$E_{n\kappa}=-M\frac{4(n+\kappa)^2-B^2}{4(n+\kappa)^2+B^2}. \tag{32}$$

Furthermore, when $n\to\infty$, one obtains $E=-M$ (continuum states), that is, it shows that when $n$ goes to infinity the energy solution of Eq. (30) becomes finite (this is the exact pspin symmetric case given by Eq. (38) of Ref. [50]).

On the other hand, to find the corresponding wave functions, referring to Eq. (29) and relations (A18) and (A22) of appendix A, we find the functions

$$\rho(s)=s^\alpha(1-s)^{2\sqrt{(\Lambda_\kappa-\frac{1}{2})^2-\tilde{\gamma}V_0}}, \tag{33}$$

$$\phi(s)=s^{\frac{\tilde{\beta}}{2\alpha}}(1-s)^{\frac{1}{2}+\sqrt{(\Lambda_\kappa-\frac{1}{2})^2-\tilde{\gamma}V_0}}. \tag{34}$$

Hence, relation (A19) gives

$$y_n(s) = P_n^{(\frac{\tilde{\beta}}{\alpha}, 2\sqrt{(\Lambda_\kappa - \frac{1}{2})^2 - \tilde{\gamma}V_0})}(1-2s). \tag{35}$$

By using $G_{n\kappa}(s) = \phi(s) y_n(s)$, we get the lower component of the Dirac spinor from relation (A24) as

$$G_{n\kappa}(s) = \tilde{B}_{n\kappa} s^{\frac{\tilde{\beta}}{2\alpha}} (1-s)^{\frac{1}{2}+\sqrt{(\Lambda_\kappa - \frac{1}{2})^2 - \tilde{\gamma}V_0}} P_n^{(\frac{\tilde{\beta}}{\alpha}, 2\sqrt{(\Lambda_\kappa - \frac{1}{2})^2 - \tilde{\gamma}V_0})}(1-2s), \tag{36}$$

where $\tilde{B}_{n\kappa}$ is normalization constant. The upper component of the Dirac spinor can be calculated from Eq. (8b) as

$$F_{n\kappa}(r) = \frac{1}{M - E_{n\kappa} + C_{ps}} \left( \frac{d}{dr} - \frac{\kappa}{r} + U(r) \right) G_{n\kappa}(r), \tag{37}$$

where $E_{n\kappa} \neq M + C_{ps}$ and with the exact pspin symmetry when $C_{ps} = 0$, only negative energy solution exists. The finiteness of our solution requires that the two-components of the wave function be defined over the whole range, $r \in (0, \infty)$. However, in the pspin limit, if the positive energy is chosen, the upper-spinor component of the wave function will be no longer defined as obviously seen in Eq. (37). Further, introducing the Coulomb-like tensor does not affect the negativity of the energy spectrum in the pspin limit, but the main contribution is just to removing the degeneracy of the spectrum.

Of course, the energy eigenvalue equation (30) admits two solutions (negative and positive), however, we choose the negative energy solution to make the wave function normalizable in the given range [50,51-53].

### 4.2. Spin symmetric case

To avoid repetition in the solution of Eq. (17), we follow the same procedures explained in the subsection 4.1 and hence obtain the following energy eigenvalue equation:

$$\left( n + \frac{1}{2} + \sqrt{(\eta_\kappa - \frac{1}{2})^2 - \gamma V_0} + \sqrt{\frac{\beta^2}{4\alpha^2}} \right)^2 = \frac{\beta^2}{4\alpha^2} - \gamma V_0, \tag{38}$$

and the corresponding wave functions for the upper Dirac spinor as

$$F_{n\kappa} = B_{n\kappa} s^{\frac{\beta}{2\alpha}} (1-s)^{\frac{1}{2}+\sqrt{(\eta_\kappa - \frac{1}{2})^2 - \gamma V_0}} P_n^{(\frac{\beta}{\alpha}, 2\sqrt{(\eta_\kappa - \frac{1}{2})^2 - \gamma V_0})}(1-2s), \tag{39}$$

where $\eta_\kappa = \kappa + H + 1$ and $B_{n\kappa}$ is the normalization constant. Finally, the lower-spinor component of the Dirac equation can be obtained via Eq. (8a) as

$$G_{n\kappa}(r) = \frac{1}{M + E_{n\kappa} - C_s}\left(\frac{d}{dr} + \frac{\kappa}{r} - U(r)\right) F_{n\kappa}(r) \tag{40}$$

where $E_{n\kappa} \neq -M + C_s$.

### 4.3. Some remarks and numerical results

The tensor potential generates a new spin-orbit centrifugal term $\Lambda(\Lambda \pm 1)$ where $\Lambda = \Lambda_\kappa$ or $\eta_\kappa$. Some numerical results are given in table 1 and table 2, where we use the parameters values as $M = 5.0\,fm^{-1}$, $V_0 = 1.0$, $C_{ps} = -5.5\,fm^{-1}$ and $C_s = 6.0\,fm^{-1}$. In table 1, we consider the same set of pspin symmetry doublets: $(1s_{1/2}, 0d_{3/2})$, $(1p_{3/2}, 0f_{5/2})$, $(1d_{5/2}, 0g_{7/2})$, $(1f_{7/2}, 0h_{9/2})$, …. Also, in table 2, we consider the same set of spin symmetry doublets: $(0p_{1/2}, 0p_{3/2})$, $(0d_{3/2}, 0d_{5/2})$, $(0f_{5/2}, 0f_{7/2})$, $(0g_{7/2}, 0g_{9/2})$, …. We see that the tensor interaction removes the degeneracy between two states in spin doublets and pspin doublets. When $H \neq 0$, the energy levels of the spin (pspin) aligned states and spin (pspin) unaligned states move in the opposite directions. For example, in pspin doublet $(1s_{1/2}, 0d_{3/2})$; when $H = 0$, $E_{1,-1} = E_{1,2} = -0.491129\,fm^{-1}$, but when $H = 5.0$, $E_{1,-1} = -0.495018\,fm^{-1}$ with $\kappa < 0$ and $E_{1,2} = -0.487533\,fm^{-1}$ with $\kappa > 0$. Also, Aydoğdu and Sever showed that the tensor interaction does not change the radial node structure of the upper and lower components of the Dirac spinor and it affects on the shape of the radial wave functions [17].

### 5. Conclusion

In this paper, under spin and pspin symmetry limits, we have obtained the approximate solutions of the Dirac equation for IQY potential by using the Nikiforov-Uvarov method. Also, we extended the exact spin and pspin symmetric solutions of the IQY potential by including the Coulomb-like tensor potential in the form of $-H/r$. Some numerical results are also given in tables 1 and 2. Obviously, the degeneracy between the members of doublet states in spin and pspin symmetries is removed by tensor interaction.

**Acknowledgments**

The authors wish to thank the kind referees for their invaluable suggestions which have greatly helped to improve this article. S. M. Ikhdair dedicates this work to the memory of his father 'Musbah' who passed away on January 26, 2012. S. M. I also acknowledges the partial support provided by the Scientific and Technological Research Council of Turkey (TÜBİTAK).

**Appendix A: Parametric Generalization of the Nikiforov-Uvarov method**

The following equation is a general form of the Schrödinger-like equation written for any potential [47]

$$\left[\frac{d^2}{ds^2} + \frac{\alpha_1 - \alpha_2 s}{s(1-\alpha_3 s)}\frac{d}{ds} + \frac{-\xi_1 s^2 + \xi_2 s - \xi_3}{[s(1-\alpha_3 s)]^2}\right]\psi_n(s) = 0. \tag{A1}$$

Comparing the above equation with Eq. (2), we get [48,49]

$$\tilde{\tau}(s) = \alpha_1 - \alpha_2 s, \tag{A2}$$

$$\sigma(s) = s(1-\alpha_3 s), \tag{A3}$$

and

$$\tilde{\sigma}(s) = -\xi_1 s^2 + \xi_2 s - \xi_3. \tag{A4}$$

Further, substituting the relations (A2)-(A4) into Eq. (7a), we find

$$\pi(s) = \alpha_4 + \alpha_5 s \pm \left[(\alpha_6 - k\alpha_3)s^2 + (\alpha_7 + k)s + \alpha_8\right]^{\frac{1}{2}}, \tag{A5}$$

where

$$\alpha_4 = \frac{1}{2}(1-\alpha_1), \tag{A6}$$

$$\alpha_5 = \frac{1}{2}(\alpha_2 - 2\alpha_3), \tag{A7}$$

$$\alpha_6 = \alpha_5^2 + \xi_1, \tag{A8}$$

$$\alpha_7 = 2\alpha_4\alpha_5 - \xi_2, \tag{A9}$$

$$\alpha_8 = \alpha_4^2 + \xi_3. \tag{A10}$$

We require that the function under square root of the relation (A5) be the square of a polynomial according to the NU method. Thus

$$k_{1,2} = -(\alpha_7 + 2\alpha_3\alpha_8) \pm 2\sqrt{\alpha_8\alpha_9}, \tag{A11}$$

where

$$\alpha_9 = \alpha_3\alpha_7 + \alpha_3^2\alpha_8 + \alpha_6. \tag{A12}$$

Therefore, for

$$k = -(\alpha_7 + 2\alpha_3\alpha_8) - 2\sqrt{\alpha_8\alpha_9}, \tag{A13}$$

we have

$$\pi(s) = \alpha_4 + \alpha_5 s - \left[\left(\sqrt{\alpha_9} + \alpha_3\sqrt{\alpha_8}\right)s - \sqrt{\alpha_8}\right]. \tag{A14}$$

Further, for the same $k$ and from Eq. (9) and the relations (A2) and (A5), we obtain

$$\tau(s) = \alpha_1 + 2\alpha_4 - (\alpha_2 - 2\alpha_5)s - 2\left[\left(\sqrt{\alpha_9} + \alpha_3\sqrt{\alpha_8}\right)s - \sqrt{\alpha_8}\right], \tag{A15}$$

and

$$\tau'(s) = -(\alpha_2 - 2\alpha_5) - 2\left(\sqrt{\alpha_9} + \alpha_3\sqrt{\alpha_8}\right)$$
$$= -2\alpha_3 - 2\left(\sqrt{\alpha_9} + \alpha_3\sqrt{\alpha_8}\right) < 0. \tag{A16}$$

When (A2) together with (A15) and (A16) are used, the following energy equation is derived:

$$\alpha_2 n - (2n+1)\alpha_5 + (2n+1)\left(\sqrt{\alpha_9} + \alpha_3\sqrt{\alpha_8}\right) + n(n-1)\alpha_3$$
$$+ \alpha_7 + 2\alpha_3\alpha_8 + 2\sqrt{\alpha_8\alpha_9} = 0. \tag{A17}$$

This equation gives the energy spectrum of the desired problem. The wave function can be calculated according to the following procedures. The weight function is obtained via Eq. (6) as

$$\rho(s) = s^{\alpha_{10}-1}(1-\alpha_3 s)^{\frac{\alpha_{11}}{\alpha_3}-\alpha_{10}-1}, \tag{A18}$$

and consequently, after the substitution in Eq. (5), we get

$$y_n(s) = P_n^{(\alpha_{10}-1,\frac{\alpha_{11}}{\alpha_3}-\alpha_{10}-1)}(1-2\alpha_3 s), \tag{A19}$$

with

$$\alpha_{10} = \alpha_1 + 2\alpha_4 + 2\sqrt{\alpha_8}, \tag{A20}$$

and

$$\alpha_{11} = \alpha_2 - 2\alpha_5 + 2\left(\sqrt{\alpha_9} + \alpha_3\sqrt{\alpha_8}\right), \tag{A21}$$

where $P_n^{(\alpha,\beta)}$ are Jacobi polynomials. Further using Eq. (4), we get the second part of wave function as

$$\phi(s) = s^{\alpha_{12}}(1-\alpha_3 s)^{-\alpha_{12}-\frac{\alpha_{13}}{\alpha_3}}. \tag{A22}$$

Hence, the total wave function becomes:

$$\psi(s) = \phi(s) y_n(s), \tag{A23}$$

$$\psi(s) = s^{\alpha_{12}}(1-\alpha_3 s)^{-\alpha_{12}-\frac{\alpha_{13}}{\alpha_3}} P_n^{(\alpha_{10}-1, \frac{\alpha_{11}}{\alpha_3}-\alpha_{10}-1)}(1-2\alpha_3 s), \ \alpha_3 \neq 0. \tag{A24}$$

Here, the constant parameters are defined by

$$\alpha_{12} = \alpha_4 + \sqrt{\alpha_8}, \tag{A25}$$

and

$$\alpha_{13} = \alpha_5 - \left(\sqrt{\alpha_9} + \alpha_3\sqrt{\alpha_8}\right). \tag{A26}$$

In some problems $\alpha_3 = 0$ [48], hence the wave functions turn into Laguerre polynomials:

$$\lim_{\alpha_3 \to 0} P_n^{(\alpha_{10}-1, \frac{\alpha_{11}}{\alpha_3}-\alpha_{10}-1)}(1-\alpha_3)s = L_n^{\alpha_{10}-1}(\alpha_{11} s), \tag{A27}$$

and

$$\lim_{\alpha_3 \to 0}(1-\alpha_3 s)^{-\alpha_{12}-\frac{\alpha_{13}}{\alpha_3}} = e^{\alpha_{13} s}. \tag{A28}$$

Therefore, the solution given in Eq. (A24) takes the form

$$\psi(s) = s^{\alpha_{12}} e^{\alpha_{13} s} L_n^{\alpha_{10}-1}(\alpha_{11} s). \tag{A29}$$

Further, in some cases, one may need a second solution of Eq. (A1). Thus, following the same procedure using the relation (A11), we obtain

$$k = -(\alpha_7 + 2\alpha_3\alpha_8) + 2\sqrt{\alpha_8\alpha_9}. \tag{A30}$$

Finally, we obtain the wave functions:

$$\psi(s) = s^{\alpha_{12}^*}(1-\alpha_3 s)^{-\alpha_{12}^*-\frac{\alpha_{13}^*}{\alpha_3}} P_n^{(\alpha_{10}^*-1, \frac{\alpha_{11}^*}{\alpha_3}-\alpha_{10}^*-1)}(1-2\alpha_3 s), \tag{A31}$$

and the energy equation:

$$\alpha_2 n - (2n-1)\alpha_5 + (2n+1)\left(\sqrt{\alpha_9} - \alpha_3\sqrt{\alpha_8}\right) + n(n-1)\alpha_3$$
$$+ \alpha_7 + 2\alpha_3\alpha_8 - 2\sqrt{\alpha_8\alpha_9} = 0. \tag{A32}$$

The pre-defined $\alpha$ parameters are:

$$\alpha_{10}^* = \alpha_1 + 2\alpha_4 - 2\sqrt{\alpha_8},$$
$$\alpha_{11}^* = \alpha_2 - 2\alpha_5 + 2\left(\sqrt{\alpha_9} - \alpha_3\sqrt{\alpha_8}\right),$$
$$\alpha_{12}^* = \alpha_4 - \sqrt{\alpha_8},$$
$$\alpha_{13}^* = \alpha_5 - \left(\sqrt{\alpha_9} - \alpha_3\sqrt{\alpha_8}\right). \tag{A33}$$

**Table 1.** The pspin symmetric bound state energy eigenvalues in unit of $fm^{-1}$ of the IQY potential for several values of $n$ and $\kappa$.

| $\tilde{l}$ | $n, \kappa < 0$ | $(l, j)$ | $E_{n,\kappa<0}$ $H=5$ | $E_{n,\kappa<0}$ $H=0$ | $n-1, \kappa > 0$ | $(l+2, j+1)$ | $E_{n-1,\kappa>0}$ $H=5$ | $E_{n-1,\kappa>0}$ $H=0$ |
|---|---|---|---|---|---|---|---|---|
| 1 | 1, -1 | $1s_{1/2}$ | −0.495018 | −0.491129 | 0, 2 | $0d_{3/2}$ | −0.487533 | −0.491129 |
| 2 | 1, -2 | $1p_{3/2}$ | −0.491129 | −0.487533 | 0, 3 | $0f_{5/2}$ | −0.484054 | −0.487533 |
| 3 | 1, -3 | $1d_{5/2}$ | −0.487533 | −0.484054 | 0, 4 | $0g_{7/2}$ | −0.480635 | −0.484054 |
| 4 | 1, -4 | $1f_{7/2}$ | −0.484054 | −0.480635 | 0, 5 | $0h_{9/2}$ | −0.477254 | −0.480635 |
| 1 | 2, -1 | $2s_{1/2}$ | −0.491152 | −0.487539 | 1, 2 | $1d_{3/2}$ | −0.484057 | −0.487539 |
| 2 | 2, -2 | $2p_{3/2}$ | −0.487539 | −0.484057 | 1, 3 | $1d_{3/2}$ | −0.480637 | −0.484057 |
| 3 | 2, -3 | $2d_{5/2}$ | −0.484057 | −0.480637 | 1, 4 | $1g_{7/2}$ | −0.477255 | −0.480637 |
| 4 | 2, -4 | $2f_{7/2}$ | −0.480637 | −0.477255 | 1, 5 | $1h_{9/2}$ | −0.473898 | −0.477255 |

**Table 2.** The spin symmetric bound state energy eigenvalues in unit of $fm^{-1}$ of the IQY potential for several values of $n$ and $\kappa$.

| $l$ | $n, \kappa < 0$ | $(l, j = l + 1/2)$ | $E_{n,\kappa<0}$ $H = 5$ | $E_{n,\kappa<0}$ $H = 0$ | $n, \kappa > 0$ | $(l, j = l - 1/2)$ | $E_{n,\kappa>0}$ $H = 5$ | $E_{n,\kappa>0}$ $H = 0$ |
|---|---|---|---|---|---|---|---|---|
| 1 | 0, -2 | $0p_{3/2}$ | 1.000000 | 0.994385 | 0, 1 | $0p_{1/2}$ | 0.990029 | 0.994385 |
| 2 | 0, -3 | $0d_{5/2}$ | 0.994385 | 0.990029 | 0, 2 | $0d_{3/2}$ | 0.985992 | 0.990029 |
| 3 | 0, -4 | $0f_{7/2}$ | 0.990029 | 0.985992 | 0, 3 | $0f_{5/2}$ | 0.982086 | 0.985992 |
| 4 | 0, -5 | $0g_{9/2}$ | 0.985992 | 0.982086 | 0, 4 | $0g_{7/2}$ | 0.978249 | 0.982086 |
| 1 | 1, -2 | $1p_{3/2}$ | 0.994367 | 0.990023 | 1, 1 | $1p_{1/2}$ | 0.985988 | 0.990023 |
| 2 | 1, -3 | $1d_{5/2}$ | 0.990023 | 0.985988 | 1, 2 | $1d_{3/2}$ | 0.982084 | 0.985988 |
| 3 | 1, -4 | $1f_{7/2}$ | 0.985988 | 0.982084 | 1, 3 | $1f_{5/2}$ | 0.978247 | 0.982084 |
| 4 | 1, -5 | $1g_{9/2}$ | 0.982084 | 0.978247 | 1, 4 | $1g_{7/2}$ | 0.974455 | 0.978247 |